%% file: main.tex
\documentclass[conference]{IEEEtran}
\IEEEoverridecommandlockouts
\usepackage{graphicx}
\usepackage[dvipsnames]{xcolor}
\usepackage{flushend}
\usepackage{amsmath}  
\usepackage{amssymb}
\usepackage{stfloats}

\usepackage{cite}
\def\BibTeX{{\rm B\kern-.05em{\sc i\kern-.025em b}\kern-.08em
T\kern-.1667em\lower.7ex\hbox{E}\kern-.125emX}}

\begin{document}

\title{Reduced-Precision Floating-Point Arithmetic in Systolic Arrays with Skewed Pipelines}

\author{
\IEEEauthorblockN{ Dionysios Filippas, Christodoulos Peltekis, Giorgos Dimitrakopoulos} 
\IEEEauthorblockA{Electrical and Computer Engineering\\ 
Democritus University of Thrace, Greece
\thanks{This work was supported by a research grant of Siemens EDA to Democritus University of Thrace for "HLS Research for Systems-on-Chip".}}
\and
\IEEEauthorblockN{Chrysostomos Nicopoulos}
\IEEEauthorblockA{Electrical and Computer Engineering\\  University of Cyprus, Cyprus}}
\maketitle

\begin{abstract}
The acceleration of deep-learning kernels in hardware relies on matrix multiplications that are executed efficiently on Systolic Arrays (SA). To effectively trade off deep-learning training/inference quality with hardware cost, SA accelerators employ reduced-precision Floating-Point (FP) arithmetic. 
In this work, we demonstrate the need for new pipeline organizations to reduce latency and improve energy efficiency of reduced-precision FP operators for the chained multiply-add operation imposed by the structure of the SA. The proposed skewed pipeline design reorganizes the pipelined operation of the FP multiply-add units to enable new forwarding paths for the exponent logic, which allow for parallel execution of the pipeline stages of consecutive PEs. As a result, the latency of the matrix multiplication operation within the SA is significantly reduced with minimal hardware cost, thereby yielding an energy reduction of 8\% and 11\% for the examined state-of-the-art CNNs.
\end{abstract}


\input{intro}
\input{baseline}

\input{proposed}

\input{evaluation}
\input{conclusions}

\bibliographystyle{IEEEtran}
\bibliography{references}

\end{document}

%% file: intro.tex
\section{Introduction}
\label{s:intro}

Deep learning has had a significant impact on many rapidly emerging applications, such as computer vision~\cite{YOLO, convnext}, natural language processing~\cite{NLP_CNN}, and robotics~\cite{CNN-SLAM}. From the outset, the widespread proliferation of various deep learning models necessitated their direct hardware acceleration, with the ultimate goal being to improve both performance and energy efficiency. 

Matrix multiplications are at the heart of deep learning algorithms and their computation in hardware maps naturally onto Systolic Arrays (SA)~\cite{why-systolic}. Tensor processing units~\cite{tpu} and other related architectures~\cite{eyriss2, scalesim, dataflow-mirroring, arrayflex} are characteristic examples of newly designed SAs.

Matrix multiplication can be implemented in SAs using integer or Floating-Point (FP) arithmetic~\cite{adaptivfloat, ten-lessons}. For increased accuracy, the use of FP arithmetic dominates during the training of deep learning models. To increase energy efficiency, inference is typically executed using integer arithmetic, after appropriate data quantization and pruning~\cite{quantization}. However, recent studies have shown that FP arithmetic cannot be avoided, if one wishes to preserve the inference quality~\cite{ten-lessons}.

\begin{figure}[th]
\centering
\includegraphics[width=0.9\columnwidth]{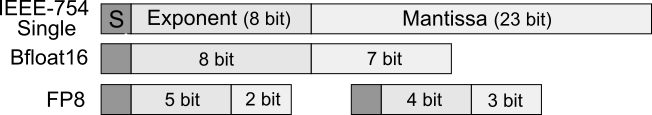}
\caption{The structure of commonly used reduced-precision FP formats.}
\label{f:fp-formats}
\end{figure}

In an effort to enjoy both benefits, i.e., the low hardware cost of integer arithmetic and the accuracy/dynamic range of FP arithmetic, several \textit{reduced-precision} FP formats have been proposed~\cite{bfloat, dlfloat, tensorfloat, NIA-fp8}.  
For instance, the 16-bit Bfloat16 format~\cite{bfloat} provides the same dynamic range as the IEEE-754 single-precision FP format, but with a smaller precision. Recently, two new 8-bit FP formats~\cite{NIA-fp8}
were proposed, which provide very similar results to those of Bfloat16, but with lower hardware cost. Fig.~\ref{f:fp-formats} illustrates these FP formats.

The introduction of reduced-precision FP formats inevitably affects the architecture of the corresponding FP operators. For instance, the operation of the traditional pipelined FP units used in SAs is dominated by the delay of the wide multipliers, while the logic dedicated to the exponent calculations is not time-critical. However, in reduced-precision FP operators this delay profile is partially flipped, since the bit-width of the mantissa (fraction) field is now equal to, or smaller than, the bit-width of the exponent field. Consequently, new architectures are required that must account for this new delay attribute of reduced-precision FP arithmetic, and, at the same time, tackle the chained structure of the SA's Processing Elements (PE). 

To address said challenges, this work proposes a novel pipeline architecture for SAs that operate on \textit{reduced-precision} FP arithmetic, with the following salient characteristics:
\begin{itemize}

\item A new \textit{skewed} pipeline micro-architecture is proposed that reorganizes the pipelined operation of the FP fused multiply-add units, thereby enabling parallel execution of the pipeline stages of consecutive PEs within the SA. The proposed design minimizes the overall latency of matrix multiplication, as compared to traditional pipelined architectures, with minimal area and power overhead.

\item Pipeline skewing is enabled by the introduction of new speculative forwarding paths within the exponent field's logic. These forwarding paths eliminate the restricting dependencies across pipeline stages and effectively increase pipeline parallelism.

\end{itemize}

Experimental evaluation using state-of-the-art CNNs demonstrates the effectiveness of the proposed architecture. The overall execution latency is markedly reduced by 16\% and 21\% for MobileNet~\cite{mobilenet} and ResNet50~\cite{resnet}, leading to overall energy reductions of 8\% and 11\% respectively. These savings were achieved with a minimal area cost of 9\%.

%% file: baseline.tex
\section{Systolic Arrays using FP Arithmetic}
\label{s:baseline}

\begin{figure}[thb]
\centering
\includegraphics[width=0.8\columnwidth]{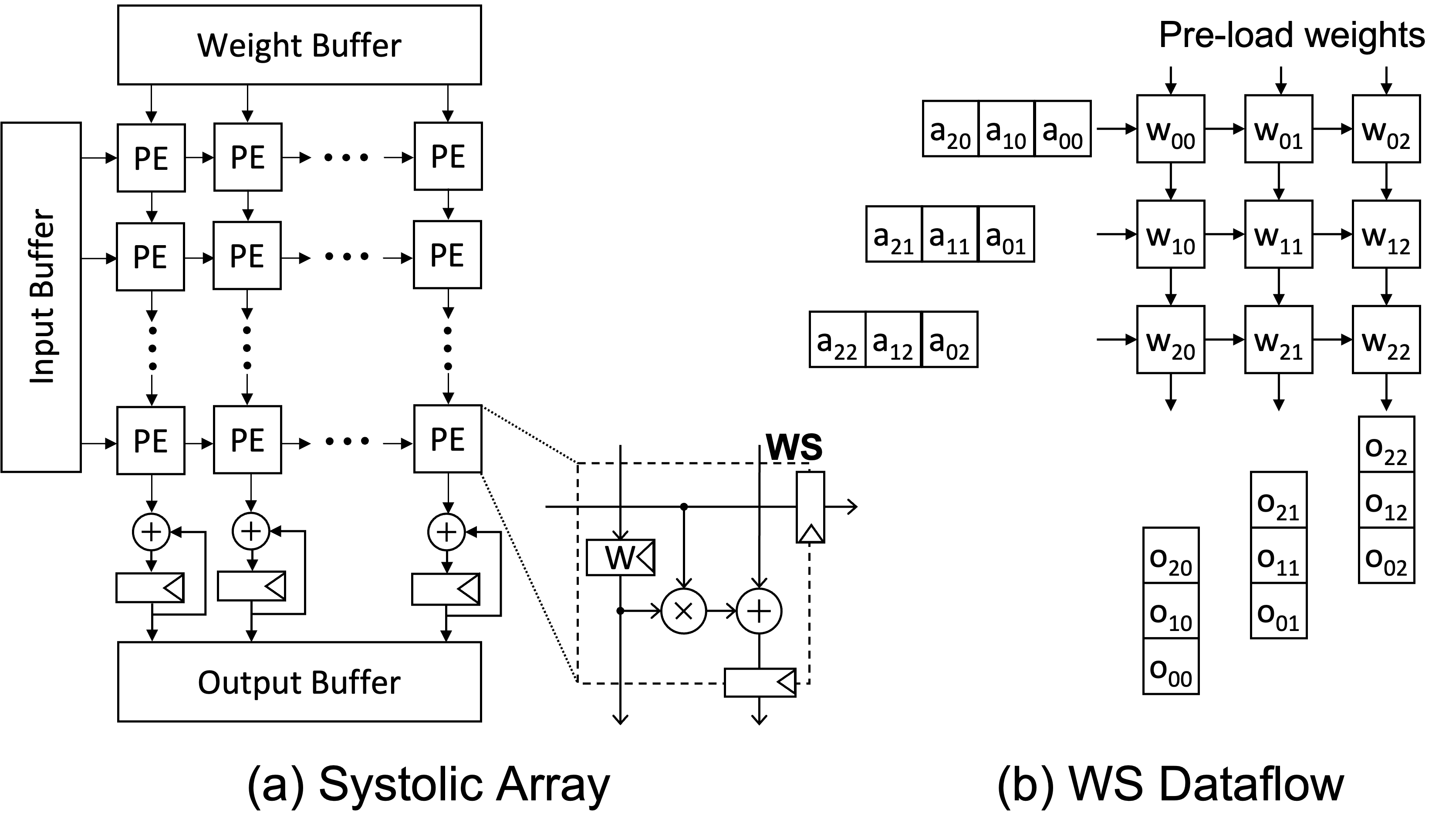}
\caption{The micro-architecture of a typical systolic array, and a high-level overview of the weight-stationary dataflow within the SA.}
\label{f:sa-baseline}
\end{figure}

The typical SA hardware structure consists of an array of PEs, as depicted in Fig.~\ref{f:sa-baseline}(a). Each PE consists of a multiplier, an adder, and necessary registers to appropriately pipeline the streaming operation. The SA is fed by local memory banks placed on the West and North edges of the array, while the output results are collected on the South edge. 

The \textit{dataflow} type employed by the SA determines the internal structure of the PEs and how the matrix multiplication $A\times W$, is executed. For instance, in \textit{weight-stationary} (WS) dataflow~\cite{scalesim}, matrix $W$ (the `weights') is pre-loaded in the SA, while matrix $A$ (the `input') is transposed and fed into the SA from the West side, as shown in Fig.~\ref{f:sa-baseline}(b). 
The WS approach is generally preferred over other dataflows, since it exploits the high spatio-temporal reuse of the weights~\cite{tpu}. After the top row is filled, it takes multiple cycles to \textit{reduce} the results of all the PEs in the same column. The number of cycles required for the reduction depends on the FP multiply-add units within each PE; i.e., the result of each PE moves downwards to the next PE in the same column. The SA becomes empty when the reduction is finished in the right-most column, for all incoming columns of matrix $A$.

Under the WS dataflow, a chain of multiply-add operations is computed in each column of the array. The FP multiply-add units in each PE have a fused/cascaded structure~\cite{galal-fma, no-fma-chain}, whereby the product of the multiplication is passed directly to the adder, without intermediate normalization and rounding. Normalization occurs after each addition at the South border of each PE. To further reduce hardware cost, state-of-the-art implementations~\cite{fused-intel, intel-nervana, fast-float} do \textit{not} perform rounding after \textit{each} multiply-add step in each PE. Instead, the rounding is performed only once, at the South end of each \textit{column}. To avoid precision loss, the intermediate results produced at the South output of each PE use double-width precision~\cite{ten-lessons}. For instance, for Bfloat16 inputs, the reduction that occurs in the vertical direction is implemented with FP32 arithmetic. 

\begin{figure}
\centering
\begin{tabular}{cc}
\includegraphics[width=0.42\columnwidth]{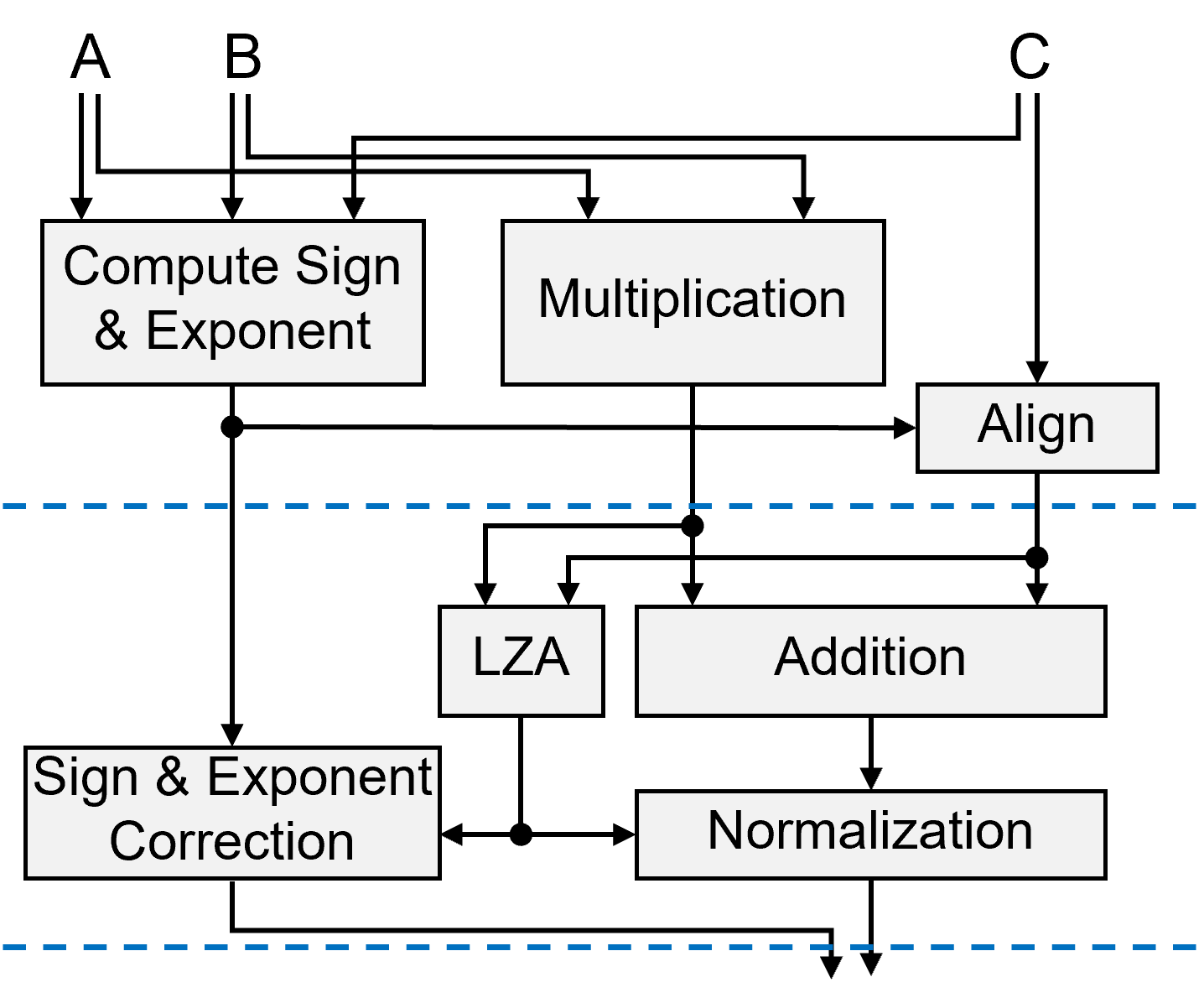}  &  
\includegraphics[width=0.42\columnwidth]{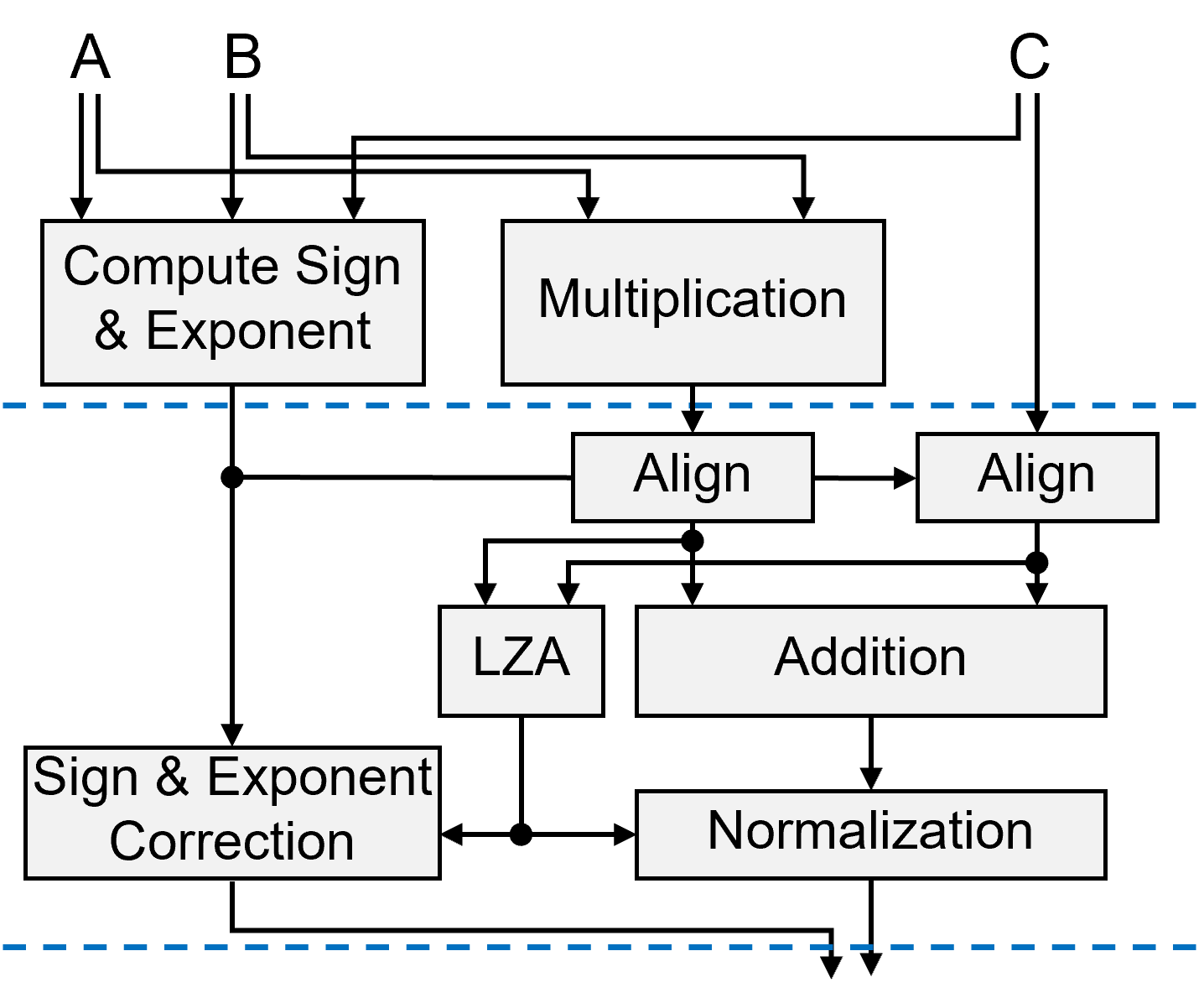}  \\
\small (a) FMA for regular precision. & \small (b) FMA for reduced precision. \\
\end{tabular}
\caption{The two main pipeline organizations that may be employed by the FP multiply-add units in each PE of the SA. In reduced-precision FP arithmetic, two pipeline stages are sufficient to achieve the required clock frequency.}
\label{f:two-stage-pipelined-fma}
\end{figure}

State-of-the-art FP multiply-add units in each PE may adopt one of the two pipelined datapaths shown in Fig.~\ref{f:two-stage-pipelined-fma}. The diagrams in the figure highlight only the most critical blocks involved in the multiply-add datapath and omit, for clarity, several logic-level details. Note that, for \textit{reduced-precision} FP arithmetic, a two-stage pipeline -- as depicted in Fig.~\ref{f:two-stage-pipelined-fma} -- is sufficient to achieve the required clock frequency. On the contrary, traditional \textit{full-precision} FP units rely on deeper pipelines for high clock frequencies~\cite{bruguera-fma-align-first}.

In the first pipeline stage of Fig.~\ref{f:two-stage-pipelined-fma}(a), multiplication is performed in parallel with the exponent computation, which calculates the amount of alignment required for the incoming partial addition result. This approach is adopted by many multiply-add architectures~\cite{bruguera-fma-align-first, multiprecision-align-first}. It is based on the fundamental assumption that the delay of the multiplier completely hides the computation on the exponents and the delay of alignment. However, this assumption is only true in full-precision FP arithmetic, where the delay of the multiplication dominates the delay of the exponent computations. 

In the second pipeline stage of Fig.~\ref{f:two-stage-pipelined-fma}(a), addition is performed. Leading-Zero Anticipation and counting (LZA)~\cite{lza, lzc}, running in parallel to the addition, predicts the amount of shifting needed to normalize the adder's result. This shift amount is also used to correct the already computed exponent of the final result. 

Since the delay of the multiplication cannot hide the delay of the exponent computations in reduced-precision FP arithmetic, it is preferable to move the alignment to the second pipeline stage, as shown in Fig.~\ref{f:two-stage-pipelined-fma}(b). The alignment may involve either the output of the multiplier, or the incoming partial addition result~\cite{fused-arm, aicas-fp}. This approach is a more natural fit to the delay profiles observed with the new FP formats. Hence, the pipeline of Fig.~\ref{f:two-stage-pipelined-fma}(b) serves as the state-of-the-art reference FP multiply-add design for reduced-precision FP arithmetic.

%% file: proposed.tex
\section{The Proposed Skewed Pipeline Architecture}
\label{s:proposed}
The two-cycle latency incurred by either the pipelined FP multiply-add units shown in Fig.~\ref{f:two-stage-pipelined-fma} increases the number of cycles required to complete the reduction within each column of the SA. Also, the pipeline parallelism across PEs is limited since the computation in each PE can begin only after the previous PE in the same column has finished its operation.

\begin{figure}[htb]
\centering
\includegraphics[width=0.95\columnwidth]{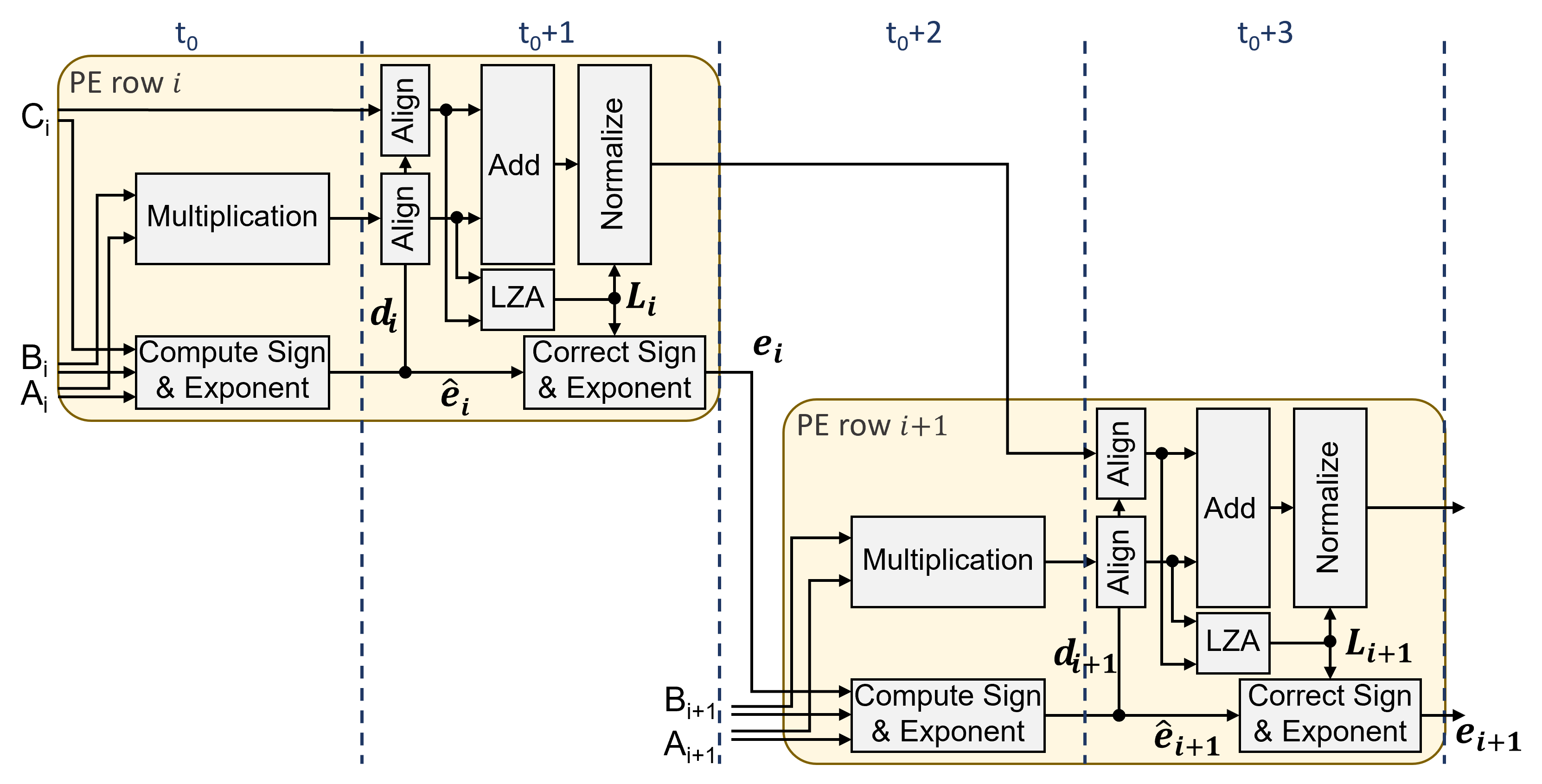}
\caption{The dependencies arising in a chained FP multiply-add operation across two neighboring PEs of the same column of the SA. These dependencies prohibit the interleaving, in time, of the pipeline execution.}
\label{f:ii-2}
\end{figure}

\subsection{The serialization problem}

The fundamental reason for this serial execution is the dependency that appears between the result of the second pipeline stage of the PE in row $i$ of the SA and the first pipeline stage of the PE in row $i+1$ of the \textit{same column}. This dependency is highlighted in Fig.~\ref{f:ii-2} across cycles $t_0+1$ and $t_0+2$. Recall that each PE employs the 2-stage pipelined organization of Fig.~\ref{f:two-stage-pipelined-fma}(b).

To increase parallelism, we would like the first pipeline stage of the PE in row $i+1$ to execute in parallel with the second pipeline stage of the previous PE (i.e., both in cycle $t_0+1$). If this were allowed, it would create a new critical combinational logic path across the two neighboring PEs, emanating from the exponent output of the first PE: the alignment logic of the first PE would be connected \textit{in series} with the LZA module of the first PE, the exponent correction logic of the first PE, and the exponent computation logic of the following PE. 

To avoid the formation of this long path, the operation in each PE begins only after the previous PE has completed its entire computation at the end of its second pipeline stage.

\subsection{Removing dependencies using speculative paths}

\begin{figure}[thb]
\centering
\includegraphics[width=0.8\columnwidth]{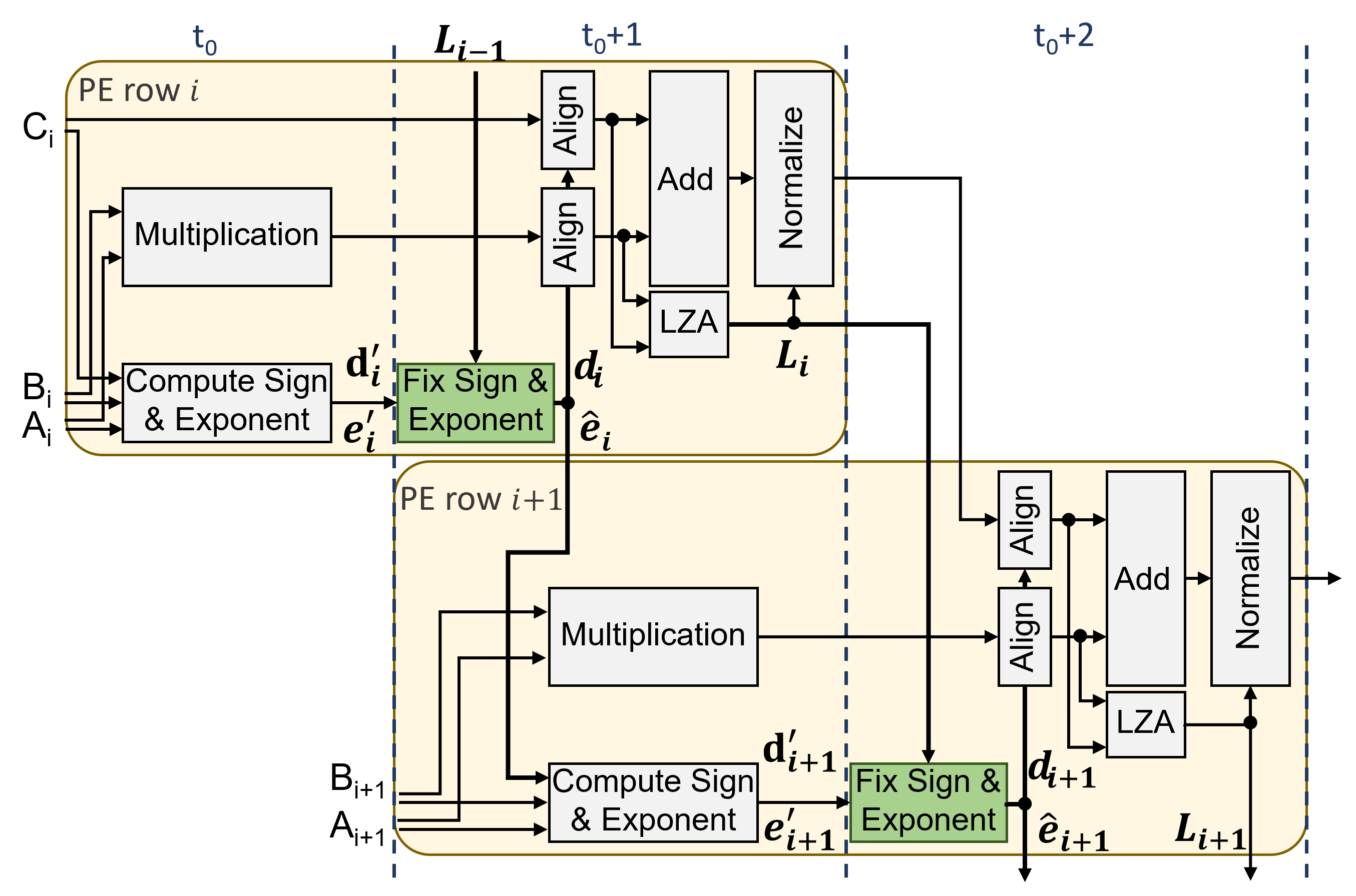}
\caption{Removing the dependency across the exponent output of each PE. A speculative exponent is produced at the output of the first pipeline stage, which is corrected at the beginning of the second stage.}
\label{f:proposed-I}
\end{figure}

To interleave, in time, the operation of the pipeline stages in each PE, a new pipelined organization for the FP multiply-add datapath is required, which relaxes the above-mentioned restricting dependencies and avoids the introduction of new combinational logic critical paths. The first step in optimizing the FP multiply-add pipeline is to decouple the exponent correction logic of the second pipeline stage of one PE from the exponent compute logic of the first pipeline stage of the next PE. This decoupling is achieved by the pipeline organization shown in Fig.~\ref{f:proposed-I}. 

In this setup, the exponent correction logic is replaced by exponent \textit{fix} logic and moved to the input of the second pipeline stage of each PE. This is the new module `Fix Sign \& Exponent' shown in green in Fig.~\ref{f:proposed-I}. To enable this relocation, the exponent fix logic no longer depends on the output of the LZA module of the current PE, but, instead, it receives the output of the LZA logic of the \emph{previous PE}. This decoupling allows for the interleaving, in time, of the pipelined execution of the multiply-add operation in consecutive PEs of the same column of the SA.

The output of the exponent fix logic controls the  alignment of the adder's inputs in the same pipeline stage and it is also given to the next PE in the place of the output exponent. This output is not the final exponent, but an intermediate and \emph{partially correct} result. The correct exponent value will be computed in the exponent fix logic of the next PE.

In each PE, the exponent compute logic selects the maximum between the exponents of the multiplication that was just calculated and the input exponent that comes from the previous PE. This maximum value, which is denoted as $\hat{e}_{i}$, represents the exponent of the \textit{unnormalized} result of the FMA's addition and it is calculated as $\hat{e}_i = max(e_{M_i}, e_{i-1})$, where $e_{M_i} = e_{A_i} + e_{B_i}$ is the exponent of the multiplication in the current PE. Furthermore, the difference of the two exponents $d_i = |e_{M_i} - e_{i-1}|$, serves as the alignment value of the two addends. In the setup of Fig.~\ref{f:ii-2}, $\hat{e}_{i}$ gets corrected by the value of the LZA $L_i$ and the corrected exponent output $e_i= \hat{e}_{i} - L_i$, now referring to the \textit{normalized} output, is forwarded to the next PE.

On the other hand, in the case of Fig.~\ref{f:proposed-I}, the compute exponent logic of the PE in row $i$ receives the intermediate $\hat{e}_{i-1}$ exponent, instead of the corrected one, as $L_{i-1}$ is not yet available to correct it. This means that the outputs of its first pipeline stage $e'_i = max(e_{M_i}, \hat{e}_{i-1})$ and $ d'_i = |e_{M_i} - \hat{e}_{i-1}|$ are \textit{speculative} values, as the exponent used refers to an unnormalized result and must be subsequently fixed. At the beginning of its second pipeline stage, $L_{i-1}$ becomes available and is forwarded to the exponent fix logic, in order to correct the speculated values. The difference of the exponents required for the alignment is:
\begin{equation*}
d_i\!=\!|e_{M_i}\!- e_{i-1}|\!=\! |e_{M_i}\!- ( \hat{e}_{i-1} - L_{i-1} )|\!=\!| (e_{M_i}\!-  \hat{e}_{i-1} ) + L_{i-1} |
\end{equation*}
As the value of $L_{i-1}$ is always greater than, or equal to, zero, we can say that:
\begin{equation*}
d_i=\begin{cases}
 |e_{M_i} -  \hat{e}_{i-1}| + L_{i-1} = d'_i + L_{i-1}\textit{ , if } e_{M_i} \ge \hat{e}_{i-1} \\
 L_{i-1} - |e_{M_i} -  \hat{e}_{i-1}| = L_{i-1} - d'_i\textit{ , if } e_{M_i} < \hat{e}_{i-1}
\end{cases}
\end{equation*}
Additionally, the fix logic generates $\hat{e}_{i}$ from $e'_i$. However, since $\hat{e}_{i}$ is either $e_{M_i}$, or $e_{i-1}$ (see above), $e'_i$ is not a computed quantity, but, instead, it comprises the two values $e_{M_i}$ and $\hat{e}_{i-1}$ that are being forwarded from the first to the second pipeline stage. After the correction of $e_{i-1} = \hat{e}_{i-1} - L_{i-1}$ in the exponent fix logic, $\hat{e}_{i}$ is computed and forwarded to the next exponent compute logic block. 

As both $\hat{e}_{i}$ and $L_i$ are computed in the same pipeline stage, and because $L_i$ becomes available at the end of the cycle, the correction of the \textit{final} exponent result (i.e., at the South edge of each column) cannot happen in the same cycle. As a result, the correction for the exponent of the last PE of each column will happen during the rounding stage at the end of the column.

The presented re-organization of the exponent computations allows for the parallel execution of the pipeline stages of consecutive PEs. However, the placement of the exponent fix logic inevitably increases the combinational path delay of the second pipeline stage of each PE. To overcome this overhead, we can \emph{retime} the normalization step. 

\begin{figure}[t!]
\centering
\includegraphics[width=0.99\columnwidth]{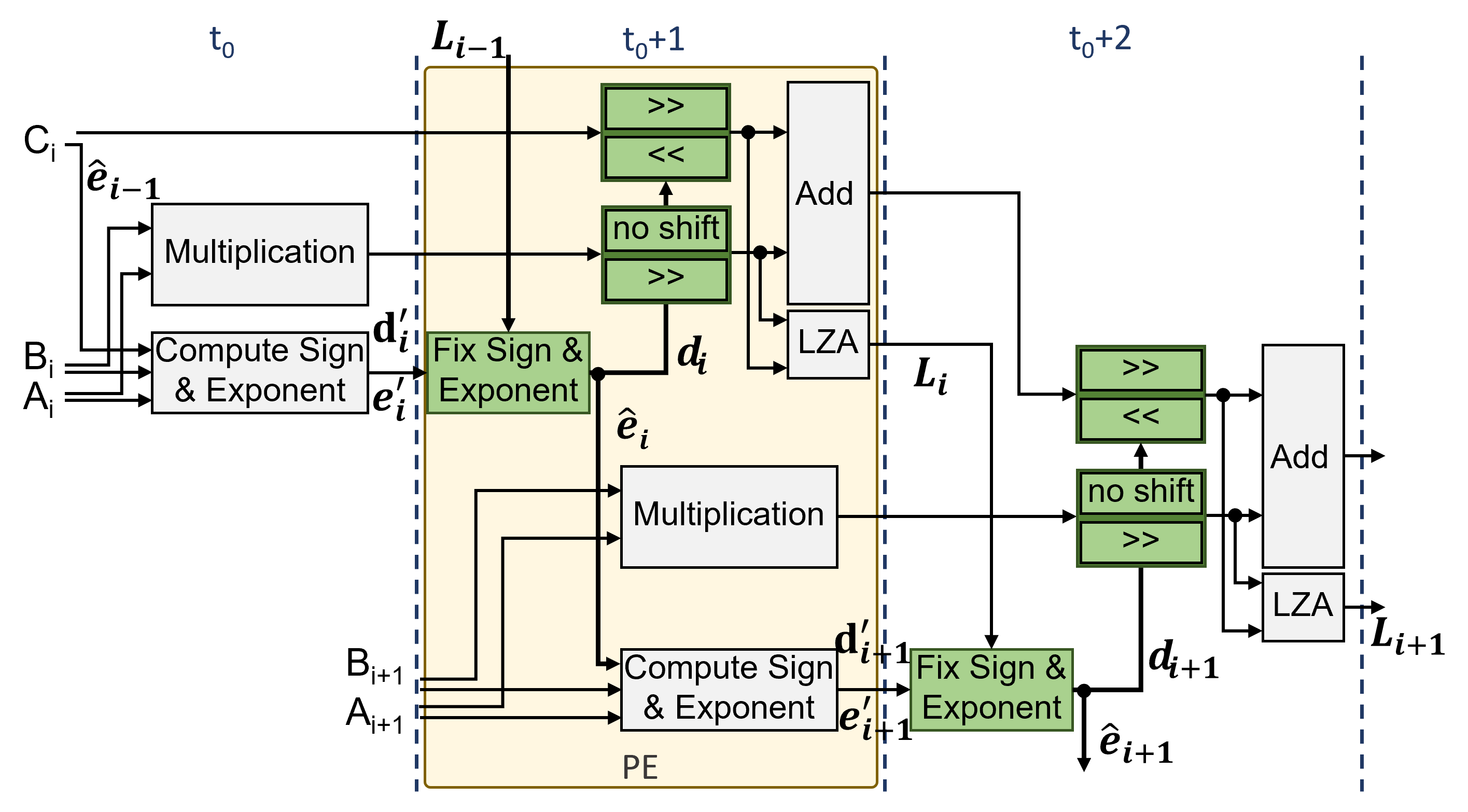}
\caption{The normalization logic is retimed in parallel to the align logic of the next PE. The addition result that flows across PEs is properly shifted to the left, or right, according to the exponent fix logic of the same stage.}
\label{f:proposed-II}
\end{figure}

This retiming is shown in Fig.~\ref{f:proposed-II}. Instead of normalizing the result of the addition in the same cycle, normalization occurs in parallel to the alignment logic at the input of the adder. The unnormalized value that arrives from the adder of the PE in row $i-1$ requires at most $L_{i-1}$ left shifts to get normalized. In the meantime, the alignment value that is computed by the fix logic determines the amount of right-shifting that may be required, if the addend was already normalized. Depending on the relation between the alignment value and $L_{i-1}$, the addend would need to either shift to the left, or to the right. As only one of these options may occur, the two operations are completely in parallel, removing the serial dependency in the delay. The new alignment scheme also affects the alignment value of the second addend that comes from the multiplication. However, only a right shift may occur in that case. The unnormalized output of the final PE will be normalized at the rounding stage at the end of the column.

\begin{figure}[t!]
\centering
\includegraphics[width=0.95\columnwidth]{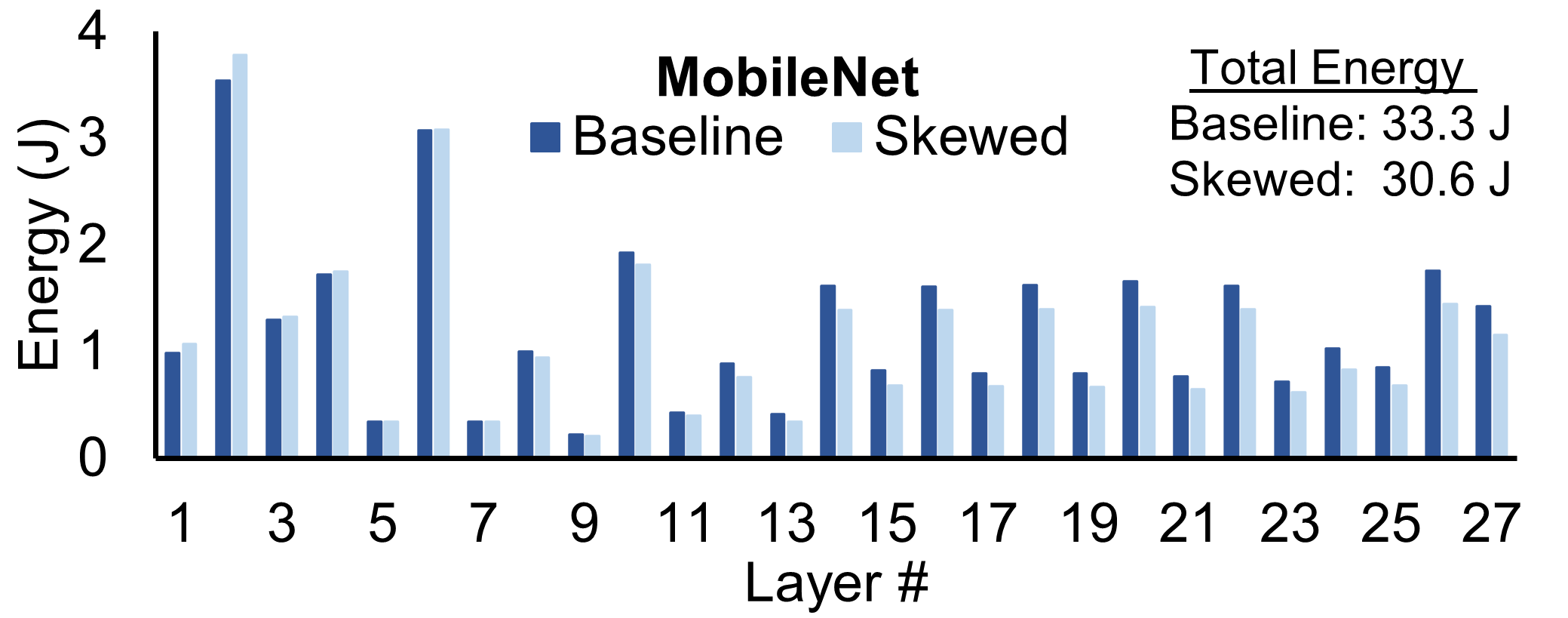}
\caption{The per-layer energy consumption when executing MobileNet~\cite{mobilenet} with the two pipeline architectures under comparison.}
\label{f:energy-spent-mobilenet}
\end{figure}

Overall, the proposed pipeline structure blurs what a PE actually is across the pipeline stages. In the new design, a PE implements, in parallel, part of the second pipeline stage of the first PE and part of the first pipeline stage of the next PE (in the same column). In fact, this new operational attribute of the PE is explicitly seen in cycle $t_0+1$ in Fig.~\ref{f:proposed-II}. Assuming that the highlighted PE of Fig.~\ref{f:proposed-II} is the last of the column, an extra addition stage is needed for the operation to be complete. Similar to the baseline case, an extra stage is also needed to round the final result of each column.

%% file: evaluation.tex
\begin{figure*}[b!]
\centering
\includegraphics[width=0.88\textwidth]{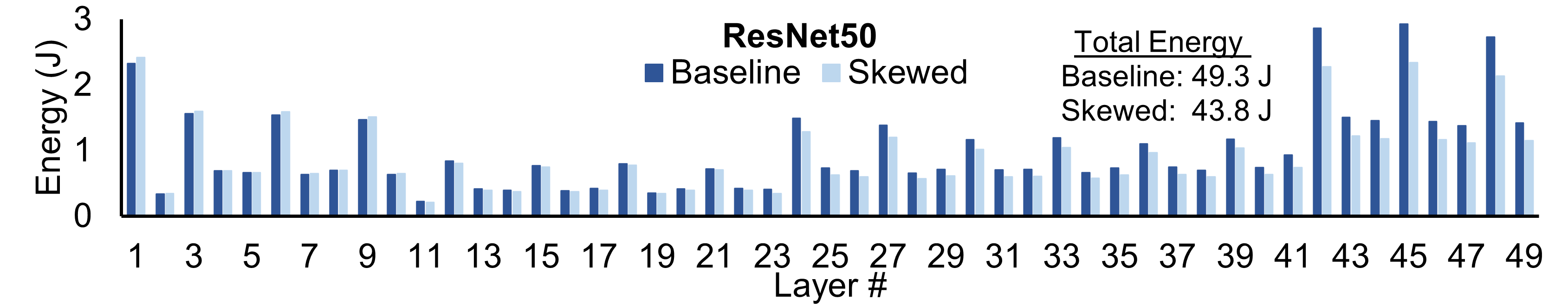}
\caption{The per-layer energy consumption when executing ResNet50~\cite{resnet} with the two pipeline architectures under comparison.}
\label{f:resnet}
\end{figure*}

\section{Evaluation}
\label{s:evaluation}

In this section, we demonstrate the effectiveness of the proposed architecture in reducing the energy requirements when computing CNNs, as compared to state-of-the-art FP multiply-add architectures employing the traditional two-stage pipeline organization of Fig.~\ref{f:two-stage-pipelined-fma}(b). In both cases, we assume Bfloat16 inputs that are reduced in the vertical direction using single precision FP32 arithmetic. 

Both designs under comparison were implemented in C++ and synthesized to Verilog RTL using Catapult HLS, driven by a commercial-grade 45-nm standard-cell library. Both SA architectures have an array size of $128\times 128$ PEs. Final timing/area results are derived from the Oasys logic synthesis tool. Power was estimated after synthesis using the PowerPro power analysis and optimization tool.

The proposed design, depicted in Fig.~\ref{f:proposed-II}, requires 9\% more area than the state-of-the-art FP multiply-add architecture shown in Fig.~\ref{f:two-stage-pipelined-fma}(b). We assume that both designs have been optimized for a clock frequency of 1 GHz. This area overhead is due to the extra pipeline registers required by the proposed design to pass intermediate exponent and LZA output values across the two pipeline stages, and the extra combinational logic of the exponent fix module. Similarly, the proposed design consumes 7\% more power, on average, when computing layers from state-of-the-art CNNs, such as 
MobileNet~\cite{mobilenet} and ResNet50~\cite{resnet}.

This marginal hardware area and power overhead is amortized by the latency savings reaped by the proposed approach, which allows for the parallel execution of the pipeline stages of consecutive PEs. Such latency savings allow the computation of each CNN layer to finish much sooner, thus yielding a \textit{reduction} in the overall \textit{energy} consumption of the computation. 

To clarify this result, Figs.~\ref{f:energy-spent-mobilenet} and~\ref{f:resnet} report the per-layer energy consumed when executing each layer of the 
MobileNet~\cite{mobilenet} and ResNet50~\cite{resnet} CNNs. The energy reported refers to the average energy observed after computing MobileNet and ResNet50 on 100 randomly picked images from the ImageNet database~\cite{imagenet}.

In both Figs.~\ref{f:energy-spent-mobilenet} and~\ref{f:resnet}, we observe that, in the first layers, the proposed approach actually leads to energy increases. The reason for this behavior is that the latency reduction cannot offset the small power overhead of the skewed pipeline organization. For the last layers, where the structure of the CNN layers changes, more latency is saved, thereby leading to significant per-layer energy savings. Most importantly, these per-layer savings translate to an \textit{overall/total} energy reduction of
8\% for MobileNet~\cite{mobilenet} and 11\% for ResNet50~\cite{resnet}.

%% file: conclusions.tex
\section{Conclusions}
\label{s:conclusions}

The design of balanced pipelined FP multiply-add units for the PEs of a SA should not stop at the boundaries of each PE, but it should also account for the dependencies arising across pipeline stages of consecutive PEs. The proposed skewed pipeline architecture focuses exactly on this aspect and effectively optimizes the latency of the reduction within each column of the SA. In effect, this reduces the overall latency of matrix multiplication. The small area and power overhead incurred by this pipeline reorganization is compensated by significant overall energy reductions when computing the layers of state-of-the-art CNNs.